\documentclass[aps,twocolumn,prb,superscriptaddress,showpacs,floatfix]{revtex4}
\usepackage{graphicx}
\usepackage{bm}
\usepackage{epsfig}

\begin{document}

\title{Hole polaron formation and migration in olivine phosphate materials}
\author{M.D. Johannes} 
\affiliation{Center for Computational Materials Science, Naval Research Laboratory, Washington, D.C. 20375}
\author{Khang Hoang}
\affiliation{Computational Materials Science Center, George Mason University, Fairfax, VA 22030}
\affiliation{Center for Computational Materials Science, Naval Research Laboratory, Washington, D.C. 20375}
\author{J.L Allen}
\affiliation{US Army Research Laboratory, 2800 Powder Mill Road, Adelphi, MD 20783}
\author{K. Gaskell}
\affiliation{Surface Analysis Center, University of Maryland, College Park, MD 20742}
\date{\today}

\begin{abstract} By combining first principles calculations and experimental XPS measurements, we investigate 
the electronic structure of potential Li-ion battery cathode materials LiMPO$_4$ (M=Mn,Fe,Co,Ni) to uncover 
the underlying mechanisms that determine small hole polaron formation and migration.  We show that small hole 
polaron formation depends on features in the electronic structure near the valence-band maximum and that, 
calculationally, these features depend on the methodology chosen for dealing with the correlated nature of 
the transition-metal $d$-derived states in these systems. Comparison with experiment reveals that a hybrid functional 
approach is superior to GGA+U in correctly reproducing the XPS spectra. Using this approach we find that 
LiNiPO$_4$ cannot support small hole polarons, but that the other three compounds can. The migration barrier 
is determined mainly by the strong or weak bonding nature of the states at the top of the valence band, 
resulting in a substantially higher barrier for LiMnPO$_4$ than for LiCoPO$_4$ or LiFePO$_4$. \end{abstract}

\maketitle 

\section{Introduction}
LiFePO$_4$ is an olivine structured material with many properties that make it attractive for usage as a cathode 
in Li-ion rechargeable batteries.  It is composed of inexpensive starting materials, and the existence of a 
stable end compound, FePO$_4$, allows for full withdrawal of Li ions.\cite{Padhi:1997p701}  However, the 
intrinsic electronic conductivity of $\sigma$ = 1.8x10$^{-8}$ S/cm is prohibitively low,\cite{Yonemura:2004p488} so that special processing techniques, such as nanostructuring, coating with carbon or 
doping with supervalent cations must be employed in order to make the material suitable for electrochemical 
cycling.\cite{Chung:2002p246, Dominko:2005p792, Delacourt:2006p727,Herle:2004p731}  The underlying mechanism by 
which these techniques are successful is still hotly debated,\cite{Ravet:2003p737,Chung:2003p739,Chung:2010p831,Hamelet:2011iz,Kang:2009p834,Zaghib:2009p880} but whatever it 
is, the conductivity can be reliably raised by a factor of 10$^6$-10$^8$, rendering LiFePO$_4$ an excellent, 
cycleable cathode material.  LiMnPO$_4$, LiCoPO$_4$, and LiNiPO$_4$ are all isostructural to LiFePO$_4$ and have 
higher voltages, both calculated and measured \cite{Padhi:1997p701,Amine:2000p711,Zhou:2004p108,Wolfenstine:2005p694}, suggesting they could be used as 
cathode materials with even more overall energy than LiFePO$_4$.  Unfortunately, these materials have similar or 
even poorer intrinsic conductivities,\cite{Oh:2010p464,Wolfenstine:2005p467,Tadanaga:2003p478,Prabu:2011p654} 
measured at $\sigma$(LiMnPO$_4$)$ < 10^{-10}$ S/cm, $\sigma$(LiCoPO$_4$)$ \sim 10^{-9}$ S/cm, and $\sigma$(LiNiPO$_4$)$ 
\sim 
10^{-9}$ S/cm. To compound the problem, techniques to raise these numbers have been only moderately successful\cite{Wolfenstine:2006p466,Kumar:2011p463,Prabu:2011p654}.  The conductivities can be raised by 10$^2$-10$^5$, 
but problems of strong capacity fade, perhaps associated with the still high resisitivity, continue to hamper 
their usefulness as practical cathode materials.

The olivine phosphates are now widely understood to be wide band-gap materials 
that exhibit polaronic rather than band-like transport.\cite{Zhou:2004p101,Zaghib:2007p392,Hoang:2011p715}  Their 
electrochemically active center is the transition-metal ion with formal valency 2$^+$ in the stoichiometric 
compound, that becomes 3$^+$ upon withdrawal of a Li ion and associated electron or when, {\it e.g.}, Li vacancies are 
created during synthesis\cite{Hoang:2011p715}.  The resulting localized hole and consequent contraction of the 
surrounding O ions are together known as a small hole polaron.  In order to move through the crystal, the small polaron must 
hop 
from one transition-metal site to another.  The barrier to this hopping creates the activated transport seen in 
experiment.\cite{Zhou:2004p101,Ellis:2006p707,Zaghib:2007p392}

In this work, we use first principles density-functional theory (DFT) and x-ray photoemission spectroscopy 
(XPS) to examine the effect of the electronic structure of LiMPO$_4$ (M=Mn,Fe,Co,Ni) on small hole polaron 
formation and migration with the goal of understanding the differences in measured conductivities and 
potentially pinpointing techniques to raise them.  We find that, calculationally, the formation of small hole 
polarons depends on which approximation to the exchange correlation functional is used.  By comparing finely 
detailed features of the density of states (DOS) in our computational results to careful XPS measurements, we 
can determine which methodology provides the best agreement with the true electronic structure.  We determine 
that the use of a hybrid functional that allows localization of electrons on both transition-metal and oxygen 
sites is necessary to reproduce the observed binding energies.  Using this functional, we calculate formation 
energies and migration barriers for small hole polarons in each of the compounds.  We find, in good agreement 
with experiment, that LiFePO$_4$ hole polarons have the highest tendency to hop, followed by LiCoPO$_4$ and 
LiMnPO$_4$.  We find that small hole polaron formation is highly unlikely in LiNiPO$_4$, which may also 
become electronically unstable with any significant delithiation.  The migration barriers depend strongly on 
the lattice distortion which, in turn, is determined by details of strong and weak bonding states formed with 
surrounding oxygen ions.

\section{Methodology} \subsection{Computational} Our calculations employ the projector augmented wave method,\cite{PAW1,PAW2} 
as implemented in the VASP code.\cite{vasp1,vasp2} For the exchange correlation potential, we use the 
Perdew-Burke-Ernzerhof version of the generalized gradient approximation (PBE-GGA),\cite{pbegga} the GGA+U 
correction\cite{Liechtenstein:1995wx} (U values were taken from the calculations of Zhou {\it et al.} 
\cite{Zhou:2004p104}), and the Heyd-Scuseria-Ernzerhof hybrid functional (HSE06).\cite{HSE061,HSE062} The 
structures are fully relaxed within each type of calculation.  The nudged elastic band (NEB)\cite{NEB} 
method was used to calculate the migration barrier of a polaron moving from one site to another.

To obtain our calculated binding energy spectra, we projected out the character of each ion from the DOS, and 
adjusted its intensity according to the Scofield cross-sections for x-rays\cite{Scofield} at 1487 eV, 
relative to the intensity of the transition-metal ion.  The adjusted partial DOS are then summed and 
convolved with a Gaussian function to approximate the temperature broadening seen in experiment. This spectra 
is compared to the experimental binding energies measured by XPS.

\subsection{Experimental}

LiMnPO$_4$ was prepared by a hydrothermal method using urea hydrolysis as the source of hydroxide ions.\cite{Allen1} MnSO$_4\cdot$H$_2$O, Li$_2$SO$_4\cdot$H$_2$O, 85\% 
H$_3$PO$_4$ and urea were dissolved in enough water to make a 20 mL clear solution. The molar ratio, Li:Mn:PO$_4$:urea, was 1:1:1:1.5. The solution was heated to 200$^{\circ}$ 
C in sealed, Teflon-lined, 45 mL autoclave for 15 hours.  The product was isolated by centrifuge and dried under vacuum at 80$^{\circ}$ C. Phase purity was determined via x-ray powder diffraction using a Rigaku Ultima III x-ray diffractometer in a parallel 
beam geometry.  Lattice constants were determined via Rietveld refinement of the parallel beam x-ray data using the Riqas program.  Lattice constants of the obtained 
LiMnPO$_4$, in spacegroup P$nma$, were $a$ = 10.456 $\AA$, $b$ = 4.7503 $\AA$, $c$ = 6.1006 $\AA$, unit cell volume = 302.99 $\AA^3$. LiFePO$_4$ was prepared similarly using 
FeSO$_4\cdot$7H$_2$O in place of MnSO$_4\cdot$H$_2$O. The lattice constants of the obtained LiFePO$_4$ were $a$ =10.3180 $\AA$, $b$=4.6901 $\AA$, $c$ = 5.9960 $\AA$, unit cell 
volume = 290.17 $\AA^3$.  LiCoPO$_4$ samples were prepared via a citrate complexation route.\cite{Allen:2011p651} Co(OH)$_2$, LiH$_2$PO$_4$, and citric acid, 1, 1.01, 1.02, 
molar ratio, respectively, were mixed into deionized water until all solids were dissolved.  The resulting solution was evaporated to dryness via a microwave oven.  The 
resulting dried mass was removed, ground lightly with mortar and pestle and heated in air at a rate of 10$^{\circ}$ C min$^{-1}$ to 600$^{\circ}$ C and the reactant mixture 
was held at this temperature for 12 h. Lattice constants were $a$ = 10.1950 $\AA$, $b$ = 5.9179 $\AA$, $c$ = 4.6972 $\AA$ for a unit cell volume of 283.40 $\AA^3$.  LiNiPO$_4$ 
was obtained from the solid state reaction of LiH$_2$PO$_4$ with Ni(OH)$_2$. The starting materials were mixed in a stoichiometric ratio via a mortar and pestle. The mixture 
was heated at 325$^{\circ}$ C for 12 hours, 500$^{\circ}$ C for 8 hours, 600$^{\circ}$ C for 12 hours and 700$^{\circ}$ C for 12 hours with intermittent grinding.  All heating 
steps were done in air.  Lattice constants were $a$ = 10.0319 $\AA$, $b$ = 5.8522 $\AA$, $c$ = 4.6779 $\AA$, unit cell volume of 274.63 $\AA^3$.

XPS spectra were acquired using a Kratos Axis 165 x-ray photoelectron spectrometer operating in hybrid mode, 
using monochromatic Al k$\alpha$ radiation (1486.6 eV) at 220 W.  The powder samples were attached to the 
sample 
holder using double sided conductive copper tape, charge neutralization was required to minimize sample 
charging.  The spectrometer was at 5 x 10$^{-8}$ Torr or lower throughout data collection. Survey spectra and Valence 
band high resolution spectra were collected with pass energies of 160 eV and 20 eV respectively. All data were 
calibrated to the hydrocarbon contamination peak at 284.8 eV.

\section{Results} \subsection{Electronic structure and hole polaron formation} 
\begin{figure}[tbph] \includegraphics[width = 0.99 \linewidth]{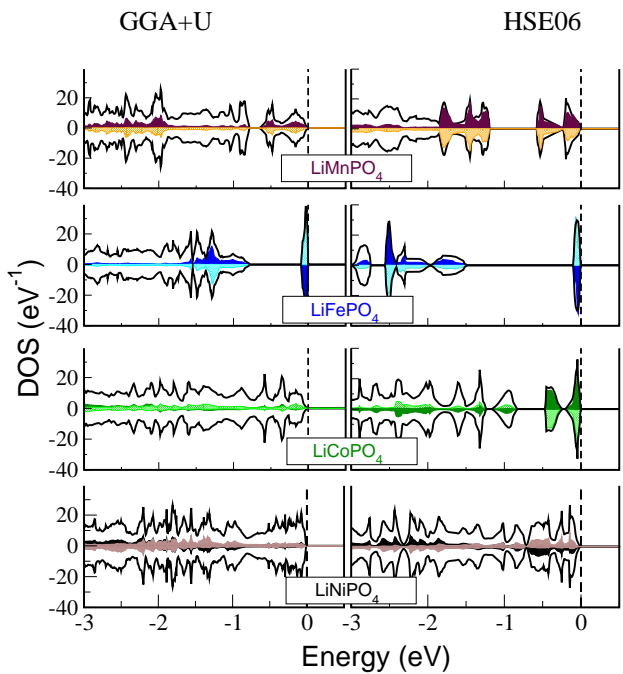} \caption{DOS plots for LiMPO$_4$ (M=Mn,Fe,Co,Ni) in the antiferromagnetic 
state.  The zero of the energy is set to the highest occupied state.  "Up" spin states are plotted on the positive y-axis and "down" spin states on the negative 
y-axis.  Shading corresponds to projected transition-metal character with majority "up" (dark) or majority "down" (light) in the antiferromagnetic ordering pattern. 
The left hand side of the Figure shows the GGA+U generated DOS, while the right side shows the HSE06 generated DOS. Considerable differences in oxygen admixture in 
the valence band are noticeable, except for M=Fe.} \label{DOS} \end{figure}

Since the creation of a hole 
polaron involves removing an electron from the valence-band maximum (VBM), the nature of the electronic states 
at and near the VBM is crucial in determining if the polaron can be formed.\cite{Hoang:2011p715} We have 
identified the following two criteria that are necessary to computationally establish small hole polarons in 
the olivine phosphate materials: (1) There must be a narrow band at the VBM that produces a sharp peak in the 
DOS, with a finite gap separating it from the broad continuum of states below it; and (2) the composition of 
this band must be predominantly transition-metal-derived, with O-derived character comprising less than 
approximately 1/3 of the total.  Furthermore, some corrective method must be applied to deal with the 
correlated transition-metal-derived $d$ states. In uncorrected GGA, these are overpenalized for localization 
and immediately delocalize throughout the crystal, whether or not the two aforementioned criteria are 
satisfied.  In Fig. \ref{DOS}, we plot the total DOS, as well as the partial DOS (shaded region) that comes 
from transition metal character (remaining character may be assumed to be oxygen) for the four olivine 
phosphate compounds.  As can be seen in Fig. \ref{DOS} and Table \ref{percent}, both criteria are filled for 
LiFePO$_4$, regardless of which exchange correlation potential is used (GGA+U or HSE06).  For LiMnPO$_4$ and 
LiCoPO$_4$, only the first criterion is filled, while the second depends on which methodology is applied.  The 
difference between the two results is significant: if oxygen character dominates at the VBM, the hole created 
upon electron withdrawal will be delocalized, whereas if transition-metal character dominates, small hole 
polarons can form and there will be a typical redox center.  Furthermore, the predominance of oxygen character 
at the VBM means that withdrawal of an electron produces oxygen ions that differ significantly from their 
preferred 2$^-$ valency, electronically destabilizing the compound and resulting in a tendency towards oxygen 
recombination and release as gas.\cite{Goodenough:2010p717,Love:2011p718} For LiNiPO$_4$ neither correction 
produces the necessary features.  In the GGA+U calculation, the Ni character has been pushed well below the 
oxygen states, leaving almost none at the VBM.  Although some Ni remains at the VBM in HSE06, the dominant 
contribution is still oxygen. Therefore, small hole polaron formation cannot be achieved.  This indicates that 
electronic transport in this compound must proceed via another mechanism, and that electronic instability 
during charging is likely.

\begin{table}[tbp] 
\caption{Percentage of transition-metal character in band nearest to the VBM in LiMPO$_4$ 
(M=Mn,Fe,Co, Ni) for three different exchange correlation schemes.  The character not attributable to the 
transition-metal at the VBM comes almost exclusively from oxygen.}
\label{table_Mper}
\begin{tabular*}{\linewidth}{@{}@{\extracolsep{\fill}} |l|ccc| @{}} \hline
& \multicolumn{3}{c|}{Transition metal character at VBM} \\ \hline
& GGA  & GGA+U & HSE06 \\
 \hline
LiMnPO$_4$   & 71\% & 39\% & 60\%  \\
LiFePO$_4$   &  88\% & 85\%  & 89\%  \\
LiCoPO$_4$        &87\% & 8\% & 78\% \\
LiNiPO$_4$        & 94\% & 9\% & 34\%   \\ \hline
\end{tabular*}
\label{percent}
\end{table}

\begin{figure*}[tbh]
\includegraphics[width=0.75 \linewidth, angle=270]{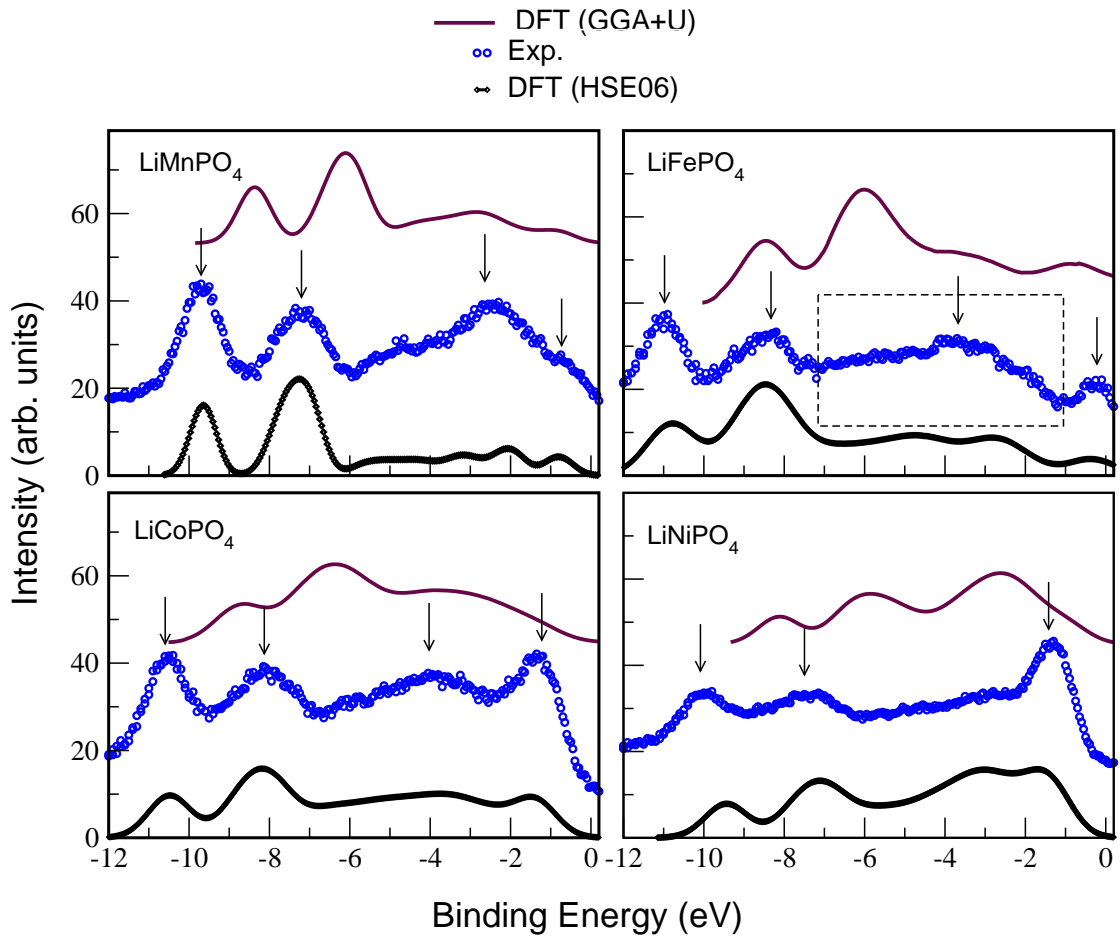}
\caption{A comparison of experimental XPS measurements of the binding energies to DFT calculations with two different
exchange-correlation
approximations.  In all cases the overall width of the spectrum and peak alignments are far better reproduced using the HSE06 methodology.}
\label{xps}
\end{figure*}

\subsection{XPS binding energies} To establish which of our methodologies, GGA+U or HSE06, better reproduces the actual electronic structure in 
the olivine phosphates, and therefore to determine whether or not small hole polaron formation is possible, we compare both to measured XPS 
spectra in Fig. \ref{xps}.  For all four compounds, the peak positions are almost perfectly captured by the HSE06 calculation, whereas the 
GGA+U spectra are significantly compressed resulting in binding energies that are too low compared to experiment.  Additionally, there are 
obvious differences near the VBM between the two methodologies and in all cases, the experimental spectra are much better matched by HSE06 than 
by GGA+U.  The spectra as a whole are better represented at both high and low binding energies using HSE06 in comparison to 
GGA+U.  

Calculated peak heights are somewhat distorted compared to measurement for all four compounds and in both methodologies. This can be
attributed to the fact that we have to project out the separate atomic characters from the DOS to apply
the Scofield cross-section parameters, and not all states are equally localized within the radii used for
projection.  This is most problematic for oxygen character which tends to be the most delocalized and
therefore most missed in the projection. Oxygen has a small experimental photoionization cross-section so
that missed character in the DOS remains uncorrected and produces exaggerated peak heights compared to
experiment.  The two peaks at highest binding energy, attributable to strongly bound P and O states, are
stronger than in experiment in all four compounds.  Since P has a high cross-section and O a low one,
small errors in the ratio of the two will cause the peaks to be artificially high.  The same effect is
operative in the LiNiPO$_4$ spectrum where Ni and O are very strongly hybridized.  The delocalized oxygen
character leads to the exaggerated peak at -3 eV.

Because an electron is withdrawn from the very top of the valence band, the character of the states there is extremely important for 
hole polaron formation.  If transition metal character dominates, charge localization is possible and hole polarons will form.  If 
oxygen character dominates, charge will delocalize and polarons will not form.  The very good match between HSE06 binding energies and 
those measured in experimental XPS, indicates that the character of these states has been well-captured by this methodology, as strong 
shifts in the character would also show up as shifts in peak positions as can be seen in the GGA+U calculations.  All spectra have two 
sharp peaks at high binding energy attributable to heavily bonded P and O states (the two leftmost arrows in Fig. \ref{xps}).  The next 
peak, located between -4 eV and -2 eV represents not a single kind of state, but a broad continumm of states composed of 
transition-metal and oxygen bonding and antibonding states which give rise to moderate secondary structure in the spectra, especially in 
in the calculations.  A peak near the valence band maximum has previously been identified as the localized state necessary for small 
hole polaron formation in LiFePO$_4$.\cite{Castro:2010p367} In all of our spectra, we see a similar low energy peak peak in all of our 
spectra, but a comparison with the unbroadened and character-resolved DOS (Fig \ref{DOS}) shows that in LiNiPO$_4$, it is {\it not} 
transition-metal dominated, nor is it even separated from the continuum of states below.  In all three other compounds, this peak 
represents a separate, energy-localized and transition-metal dominated state in which polaron formation is possible.  Temperature 
broadening mainly obscures the separation between this "polaron state" and the continuum, but close inspection of the experimental 
spectrum does allow the interpretation that LiNiPO$_4$ has only three clearly distinguishable peaks (plus a small shoulder) due to a merged first and second 
peak, whereas the other compounds have four.  Comparison with calculation makes this interpretation the most reasonable.  The lowest 
energy peak in LiNiPO$_4$ is more closely associated with the second lowest peak in the other compounds and is not suitable as a "polaron 
state". We conclude that the VBM of LiNiPO$_4$ is dominated by oxygen character, that no gap between the initial peak and the rest of 
spectrum exists and thus, small hole polarons will not form during delithiation of LiNiPO$_4$ and oxygen will likely be evolved.

\begin{figure}[tbh]
\includegraphics[width=0.95 \linewidth]{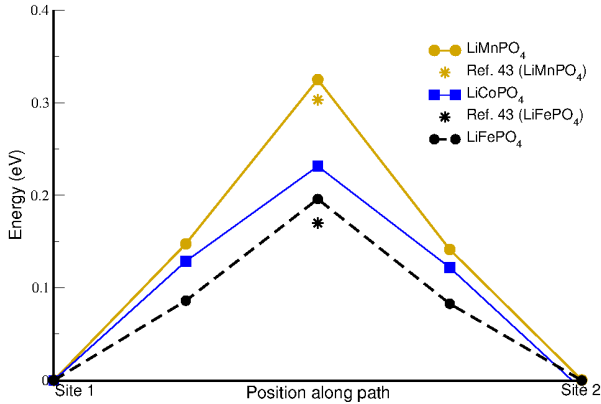}
\caption{Energy of polaron migration as calculated at steps along a linearly interpolated path between two calculated polarons.  Comparison with a previous calculations is provided for reference.}
\label{migration}
\end{figure}

\subsection{Small hole polaron conduction} To understand how small hole polarons move through the olivine structure, we used a 
method 
similar to that of Refs. \onlinecite{Maxisch:2006p103,Ong:2011he} to calculate the migration barrier for a polaron 
hopping from one transition metal site to another.  First we remove an electron from a given transition-metal site in a 
sixteen formula unit supercell of LiMPO$_4$.  We fully relax all the ionic positions to achieve 
the oxygen distortion around the hole, along with any other nearby distortions that occur.  In a separate 
calculation, we remove an electron from an adjacent transition-metal site and again perform a relaxation.  We 
find that the distortion of the lattice is negligible beyond a single unit cell, and is mainly constrained to 
the first shell of neighbors surrounding the hole, confirming the small polaron designation.  We use a linear 
interpolation of the two structures and calculate the energies along a series of intermediate positions, using 
the NEB method.  The highest energy we consider to be the polaron migration barrier.  The positive (hole) charge 
density in this scenario "follows" the distortion, hopping wholly or partially from one transition-metal site to the next without 
ever occupying an intermediate space. The migration barrier, therefore reflects mainly the energy required to 
distort the lattice. As can be seen in Fig. \ref{migration}, the resulting numbers are 0.33 eV for LiMnPO$_4$, 0.20 eV for LiFePO$_4$, and 0.23 eV for 
LiCoPO$_4$, following the trend of measured conductivities in these materials.  The numbers for LiMnPO$_4$ and 
LiFePO$_4$ are shifted slightly higher, but otherwise in good agreement with Ref. \onlinecite{Ong:2011he} using the same exchange correlation 
potential.

\section{Discussion}

\subsection{Effect of exchange-correlation corrections} In a Hartree-Fock calculation, the subtraction of 
the Hartree and exchange terms produces an exact cancelation of the electron self-interaction energy.  
When an approximation to the exchange term is made in DFT, this exact cancellation is compromised and 
leftover self-interaction contributes to the eigenvalues and total energy of the system.\cite{Parr:1989wg} 
This is especially problematic for systems with localized bands.  Both the GGA+U and HSE06 methods are 
useful to counteract the erroneous self-interaction term, but they are different in their effects on the 
electronic structure.  The GGA+U method applies a correction to the correlated bands (transition-metal 
$d$-bands for the purposes of this article) only.  Since all non-correlated bands are unaffected, one 
effect is that the occupied (down-shifted) $d$ bands are brought closer in energy to the oxygen bands 
below them, compared to GGA alone.\cite{VIA91} The hybridization between the O $p$ bands and metal $d$ 
bands depends inversely on the energy distance between them, which systematically decreases with the 
increasing $Z$-value of the transition-metal ion.  Consequently, the smaller energy separation in GGA+U 
causes a large increase in the amount of oxygen mixed into the transition-metal-derived bands at the 
valence band maximum (See Table \ref{table_Mper} and Fig. \ref{DOS}).  For LiCoPO$_4$ and LiNiPO$_4$, 
where the $d$ states are already quite low in energy, the downshift is so dramatic that the $d$-derived 
bands are located within or even below the oxygen band complex resulting in a {\it predominantly} oxygen 
VBM character, which does not accurately represent the experimental data.

The HSE06 method, on the other hand, incorporates exact exchange along with the GGA 
approximation to the exchange-correlation potential, with each being used in a certain region of real 
space.\cite{HSE061}  The exact exchange portion does not require any discrimination between "correlated" and 
"non-correlated" orbitals and therefore shifts both metal-derived and O-derived states downward, though not 
equally.  Compared to GGA+U, the mixture of O character into the highest energy valence states is greatly 
reduced.  Comparison with the XPS spectra show that this reduced hybridization better reproduces the 
experimental binding energies. For all but LiNiPO$_4$, this results in a sharp peak with mainly 
transition-metal character at the VBM and a gap to the continuum of states below it.  For the Ni-based 
compound, oxygen character at the VBM is reduced from the GGA+U value, but is still too high to allow small 
hole polaron formation. The good match between experiment and calculation shows that this is due to 
hybridization resulting from the legitimately low energy position of the Ni $d$ bands, and {\it not} due to 
computational artifacts.

\begin{table}[tbp]

\caption{Comparison of M-oxygen bond lengths for LiMPO$_4$ (M=Mn,Fe,Co) at M sites with a localized hole 
polaron (M$^{3+}$) and without a hole polaron (M$^{2+}$), obtained from supercell calculations.  
Contraction around the hole occurs in all cases, although a Jahn-Teller distortion expands two bonds in 
LiMnPO$_4$).}

\begin{tabular}{|l|cccccc|c|}
\hline
LiMnPO$_4$& \multicolumn{6}{c|}{M-O bond lengths} & Ave \\
 \hline
Mn$^{2+}$ & 2.13 & 2.13 & 2.15 & 2.25  &  2.27  &  2.27 & 2.20  \\
Mn$^{3+}$  & 1.93 & 1.95 & 2.00 & 2.00  &  2.34  &  2.34 & 2.09  \\ \hline
LiFePO$_4$& &&&&&&  \\
 \hline
Fe$^{2+}$ & 2.06 & 2.06 & 2.12 & 2.21  &  2.25  &  2.25 & 2.16  \\
Fe$^{3+}$  & 2.01 & 2.01 & 2.02 & 2.07  &  2.13  &  2.14 & 2.06  \\ \hline
LiCoPO$_4$& &&&&& & \\
 \hline
Co$^{2+}$ & 2.04 & 2.05 & 2.05 & 2.11  &  2.16  &  2.18 & 2.10  \\
Co$^{3+}$  & 1.94 & 1.94 & 2.00 & 2.10  &  2.13  &  2.14 & 2.04  \\ \hline
\end{tabular}   
\label{contract}
\end{table}

\subsection{Lattice changes during polaron formation} 
\begin{figure}[tbhp] \begin{tabular}{ccc} \epsfig{file=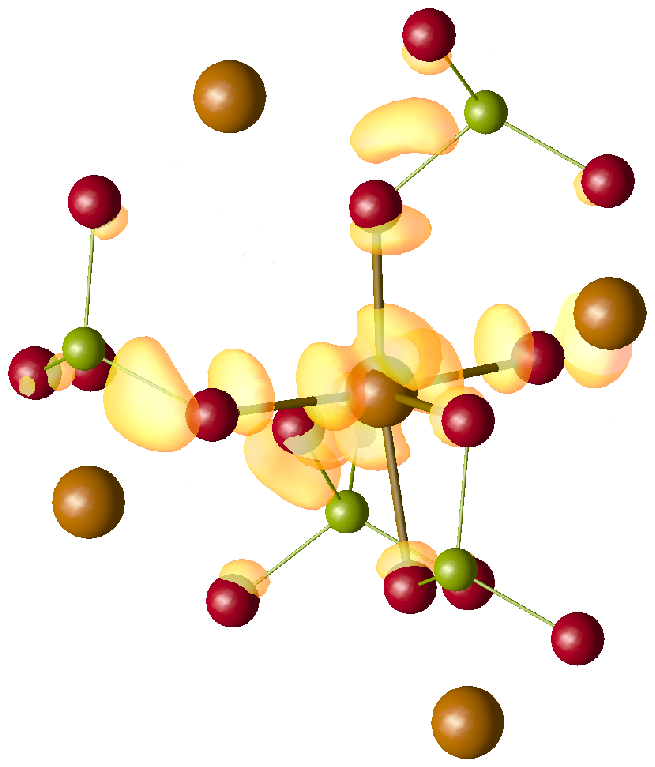, width = 1.0in}& \epsfig{file=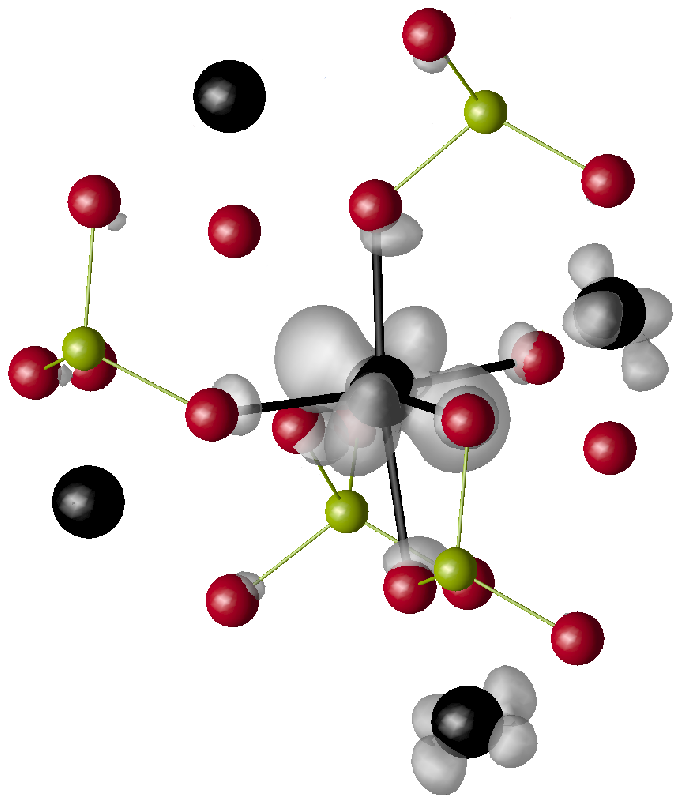, width = 1.0in}& 
\epsfig{file=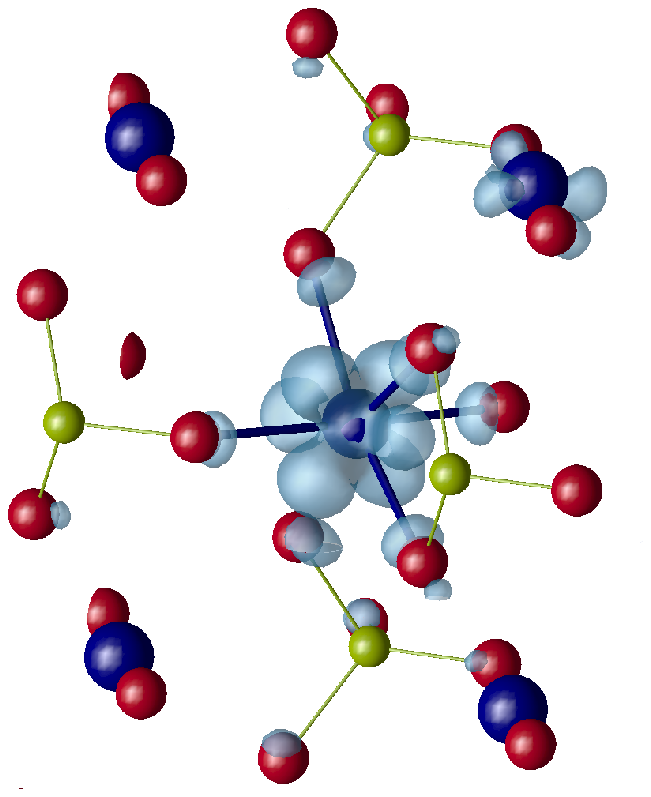, width = 1.0in}\\ \epsfig{file=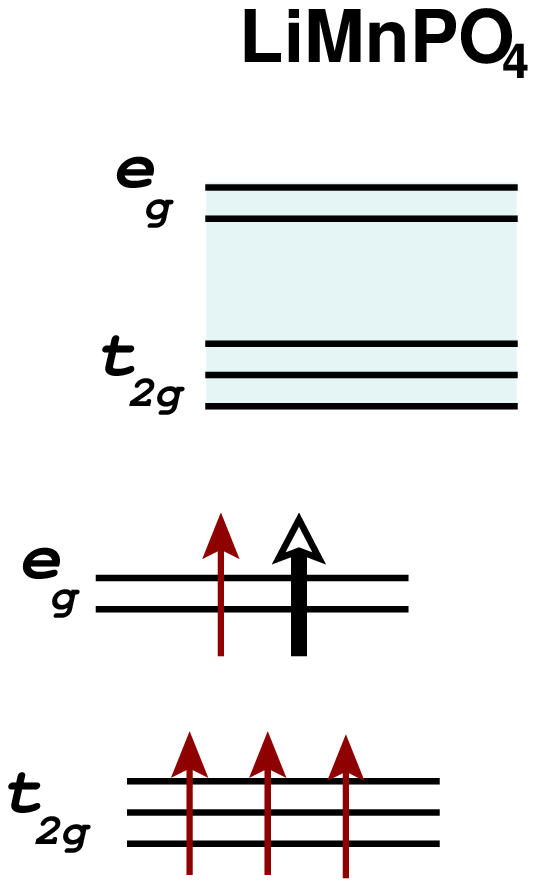, width = 1.0in}& \epsfig{file=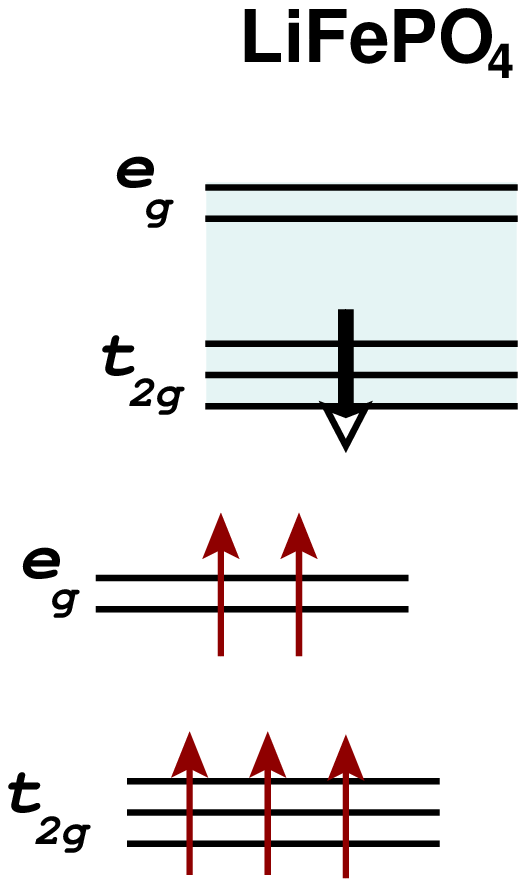, width = 1.0in}& 
\epsfig{file=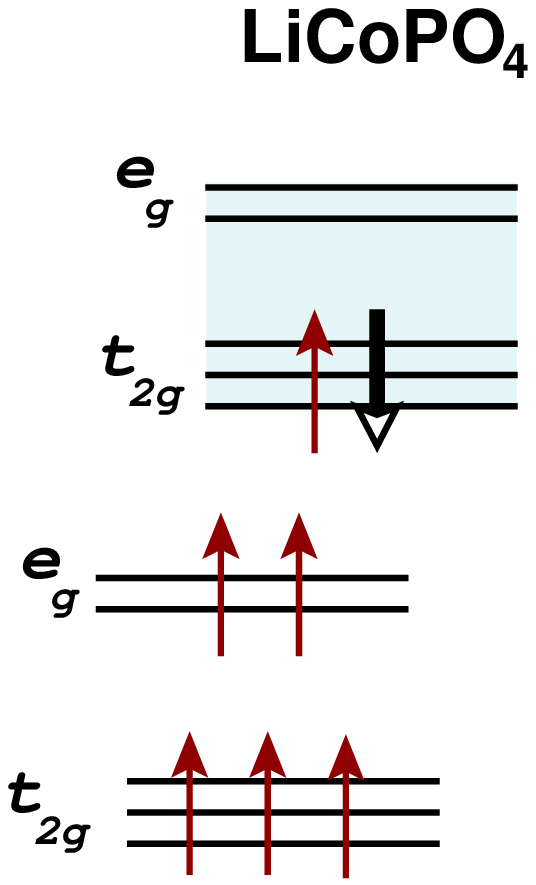, width = 1.0in} \end{tabular} \caption{Charge density of the small hole polaron in phosphate olivine 
compounds.  Most of the positive (hole) charge resides on the transition-metal, but some charge is also on surrounding oxygens, 
particularly in the case of LiMnPO$_4$ for which the hole sits in a strong bonding, heavily oxygen-hybridized state. Below each 
polaron density plot is a schematic showing the energy levels and fillings in the (approximately) octahedral environment of each 
compound.  Strong-bonding $e_g$ and weak-bonding $t_{2g}$ states are show with the majority spin states (left) widely exchange split from the shaded minority spin states (right). The heavy, hollow 
arrow indicates the position of the removed electron or hole polaron.} \label{pols} \end{figure}

The lattice contraction around the removed electron depends strongly on the occupation of 
the metal $d$ states prior to withdrawal.  In the olivine structure, the transition metal sits in a quasi-octahedral MO$_6$ environment.  The 
crystal field and ligand interactions split the 5 $d$ states into a lower three-fold degenerate manifold ($t_{2g}$) and an upper doubly 
degenerate manifold ($e_g$).  The states are strongly exchange split such that the ground state is high spin in all cases.  For LiMnPO$_4$, the 
hole state is in a $e_g$ state, while for LiFePO$_4$ and LiCoPO$_4$, the hole sits in a $t_{2g}$ state (See Fig. \ref{pols}).  The anti-bonding 
$e_g$ states are heavily mixed with oxygen - the result of strong $\sigma$-type bonding between metal and oxygen states, whereas the weak- or 
non-bonding $t_{2g}$ states are much more lightly mixed with oxygen.  Removing an electron from the anti-bonding states increases the bond 
strength and the surrounding octahedron contracts strongly.  Removing an electron from the non-bonding states results in a much weaker shift of 
oxygen toward the transition-metal ion.  This can be seen in Table \ref{contract} where the average Mn-O bond distance changes much more than 
Fe-O or Co-O.  Lattice distortion around the Mn$^{3+}$ ion is complicated by the fact that creation of a hole leaves a single electron in the 
doubly degenerate $e_g$ complex, stimulating a Jahn-Teller distortion which partially counteracts the overall contraction of the MnO$_6$ 
octahedra.  The two-long/four-short pattern typical of octahedral Jahn-Teller distortions is combined with an overall shrinking of all Mn-O 
bonds to produce four dramatically shorter bonds and two slightly longer ones.  Thus, the average contraction in bond length (5\%) belies the 
actual magnitude of the distortion: a 9\% contraction of four bonds and a 3\% expansion of two others.  The average contraction of bond lengths 
in LiFePO$_4$ is 4.3\% and in LiCoPO$_4$ is 2.7\%.  The strong distortion in the Mn-based material will disrupt small hole polaron migration 
(electronic conduction) since large local changes must propagate through the crystal.  Such distortions are also likely to be detrimental to 
the structure during repeated electrochemical cycling, especially at high rates.

\section{Summary} Our investigation of four olivine phosphate compounds reveals that small hole polaron 
formation depends on the existence of a narrow, isolated band at the VBM that contains predominantly 
transition-metal character.  The relevant electronic structure is sensitive to the particular 
approximation used for the exchange-correlation potential.  GGA+U pushes the transition-metal bands down 
much further relative to the oxygen bands than does HSE06, thereby increasing the hybridization and mixing 
more oxygen into the valence band.  A comparison with experimental XPS spectra shows that only HSE06 
properly captures the true electronic structure for these materials at the VBM. Using this methodology, we 
find that LiMnPO$_4$, LiFePO$_4$, and LiCoPO$_4$ can all support small hole polarons, while LiNiPO$_4$ 
cannot due to a predominantly oxygen-derived valence band that delocalizes the hole upon electron removal. 
The hole polaron state in LiMnPO$_4$ is within the strong-bonding $e_g$ complex and additionally supports 
a Jahn-Teller distortion, resulting in an overall stronger local deformation of the lattice and therefore 
higher migration barrier than in LiFePO$_4$ or LiCoPO$_4$ where the hole polaron occupies a weak- or 
non-bonding state.  The order of barriers is LiFePO$_4 <$ LiCoPO$_4 <$ LiMnPO$_4$, in agreement with 
experimental measurements of conductivity.

\acknowledgments 
We acknowledge useful technical discussions with Jeff Wolfenstine, T. Richard Jow and 
Shyue-Ping Ong.  Funding was provided by the Army Research Laboratory and the Office of Naval Research.

\end{document}